\documentclass{PoS}
\usepackage{bm,bbm,amssymb,graphicx}
\def\sl{\!\!\!\!\!/}

\title{\vspace{-2cm}
\rightline{\normalsize\rm ANL-HEP-CP-13-4}
\vspace{1.3cm}
Endpoint Logarithms in 
$\bm{e}^{\bm{+}}\bm{e}^{\bm{-}}\bm{\to J/\psi+\eta_c}$
}

\ShortTitle{Endpoint Logarithms in $e^+e^- \to J/\psi + \eta_c$}

\author{Geoffrey T. Bodwin%
\thanks{Work supported by
the U.~S.~Department of Energy, Division of High Energy Physics, under Contract
No.~DE-AC02-06CH11357}\\
        HEP Division, Argonne National Laboratory\\
        E-mail: \email{gtb@hep.anl.gov}}

\author{\speaker{Hee Sok Chung}%
        \\
        Department of Physics, Korea University\\
        E-mail: \email{neville@korea.ac.kr}}

\author{Jungil Lee\\
        Department of Physics, Korea University\\
        E-mail: \email{jungil@korea.ac.kr}}

\abstract{
We investigate the origin of the double logarithms of $Q^2/m_c^2$ that
appear in the calculation of the cross section for $e^+e^-\to J/\psi +
\eta_c$ at next-to-leading order in the strong coupling $\alpha_s$.
Here, $Q^2$ is the square of the center-of-momentum energy, and $m_c$
is the charm-quark mass. We find that, diagram-by-diagram, the double
logarithms are accounted for by Sudakov double logarithms and endpoint
double logarithms. The Sudakov double logarithms cancel in the sum over
all diagrams, but the endpoint double logarithms do not. We reinterpret
the endpoint double logarithms in terms of a leading region of loop
integration in which a spectator fermion line becomes soft and
collinear. This reinterpretation may simplify the process of
establishing an all-orders factorization theorem for this helicity-flip
process, which, in turn, might allow one to resum logarithms of
$Q^2/m_c^2$ to all orders in $\alpha_s$. }

\FullConference{Xth Quark Confinement and the Hadron Spectrum\\
                 8-12 October 2012\\
                 TUM Campus Garching, Munich, Germany}
\begin{document} 

\section{Introduction}

The exclusive double-charmonium production cross section $\sigma (e^+
e^- \to J/\psi + \eta_c)$, which has been measured by the Belle
\cite{Abe:2002rb, Abe:2004ww} and {\it BABAR} \cite{Aubert:2005tj}
collaborations, has provided serious challenges to the nonrelativistic
QCD (NRQCD) factorization formalism
\cite{Bodwin:1994jh,Bodwin:2008nf,Bodwin:2009cb,Bodwin:2010fi,Bodwin:2010xb}.
The theoretical value for the cross section at leading order (LO) in the
strong coupling $\alpha_s$ and the heavy-quark velocity
$v$~\cite{Braaten:2002fi, Liu:2002wq} is almost an order of magnitude
smaller than the measured rate. The discrepancy between theory and
experiment seems to have been resolved by the inclusion of higher-order
corrections, which include QCD corrections of next-to-leading order
(NLO) in $\alpha_s$~\cite{Zhang:2005cha,Gong:2007db} and
corrections of higher order in $v$ (relativistic corrections)
\cite{Bodwin:2006dn,Bodwin:2006ke,Bodwin:2007ga}. Here, $v$ is the
velocity of the heavy quark ($Q$) or heavy antiquark ($\bar Q$) in the
quarkonium rest frame.

As the authors of Ref.~\cite{Jia:2010fw} have pointed out, the large NLO
QCD corrections involve double logarithms of $Q^2/m_c^2$, where $Q^2$ is
the square of the $e^+e^-$-center-of-momentum energy and $m_c$ is the
charm-quark mass. These logarithms are sufficiently large that it may be
necessary to resum them to all orders in $\alpha_s$ in order to obtain a
reliable theoretical prediction. Typically, the resummation of large
logarithms is carried out through a factorization of the contributions
that arise from small momentum scales from the contributions that arise
from large momentum scales. A first step in deriving such a
factorization is to identify the regions of loop-momentum integrals that
produce the large logarithms at the leading nonzero power of
$Q^2$~\cite{Collins:1984kg}.

In this paper, we identify the leading loop-momentum regions that give
rise to the double logarithms in the NLO QCD corrections to the process
$e^+ e^- \to J/\psi + \eta_c$. According to the NRQCD factorization
formalism~\cite{Bodwin:1994jh}, the amplitude for this process at LO in
$v$, can be written as a product of a short-distance coefficient with the
NRQCD long-distance matrix elements (LDMEs) for the evolution of $Q\bar
Q$ pairs into the $J/\psi$ and the $\eta_c$. The same short-distance
coefficient appears in the amplitude for the production of two $Q\bar Q$
pairs that have the same quantum numbers as the corresponding quarkonia.
The short-distance coefficient can be obtained perturbatively by
comparing the full-QCD amplitude $i {\cal A} [e^+ e^- \to Q \bar Q_1
({}^3S_1) + Q \bar Q_1 ({}^1S_0)]$ with the NRQCD amplitude, which
consists of the short-distance coefficient times the NRQCD LDMEs for the
$Q\bar Q$ states. (Here the subscripts $1$ indicate that the $Q\bar Q$
pairs are in color-singlet states.) The double logarithms of $Q^2/m_c^2$
arise solely from the full-QCD amplitude because the NRQCD LDMEs for the
$Q \bar Q$ states are insensitive to momentum scales of order $m_c$ or
larger. In Sec.~\ref{sec:double_logarithms}, we carry out the
calculation of the double logarithms of $Q^2/m_c^2$ at NLO in $\alpha_s$
for the process $e^+ e^- \to J/\psi+\eta_c$ by examining the NLO QCD
corrections to the full-QCD process. We find that the double logarithms
arise from Sudakov or endpoint regions of loop momenta, and we identify
the contribution that arises from each region for each NLO Feynman
diagram. Our results for the double logarithms agree with those that
were obtained in the complete NLO calculations
\cite{Zhang:2005cha,Gong:2007db}. We find that the Sudakov double
logarithms cancel in the sum over diagrams and that power-divergent
contributions from the endpoint region vanish. In
Sec.~\ref{sec:sudakov}, we give a general analysis of the Sudakov double
logarithms that elucidates the reason for their cancellation in the sum
over diagrams. In Sec.~\ref{sec:endpoint}, we give a general analysis of
the endpoint region that establishes the absence of power-divergent
contributions. We summarize our results in Sec.~\ref{sec:summary}.

\section{Calculation of double logarithms}
\label{sec:double_logarithms}%

In this section, we evaluate the double logarithms that appear in the NLO
QCD corrections to the amplitude $e^+ e^- \to J/\psi+\eta_c$, and we
identify the momentum regions that are associated with the logarithms.
The process $e^+ e^- \to Q \bar Q_1 ({}^3S_1) + Q \bar Q_1 ({}^1S_0)$ is
composed of $e^+ e^- \to \gamma^*$, followed by $\gamma^* \to Q \bar Q_1
({}^3S_1) + Q \bar Q_1 ({}^1S_0)$. Because the process $e^+ e^- \to
\gamma^*$ does not receive QCD corrections in relative order
$\alpha^0\alpha_s$, we need to consider only the process $\gamma^* \to
Q \bar Q_1 ({}^3S_1) + Q \bar Q_1 ({}^1S_0)$ in the evaluation of the
NLO QCD corrections.

\subsection{Kinematics, conventions, and nomenclature}

Now we describe the kinematics, conventions, and nomenclature that we
use in calculating the double logarithms and throughout this paper. We
work in the Feynman gauge. We use the light-cone momentum
coordinates $k=[k^+,k^-,\bm{k}_\perp]=
[(k^0+k^3)/\sqrt{2},(k^0-k^3)/\sqrt{2},\bm{k}_\perp]$ and work in the
$e^+e^-$-center-of-momentum frame. Because our calculation is at
LO in $v$, we set the relative momentum of the $Q$ and $\bar Q$ in each
charmonium equal to zero. Then, the momenta of the $Q$ and $\bar{Q}$ in
the $J/\psi$ are both
$p=[(\sqrt{P^2+m_c^2}+P)/\sqrt{2},(\sqrt{P^2+m_c^2}-P)/\sqrt{2},\bm{0}_\perp]$,
and the momenta of the $Q$ and $\bar{Q}$ in the $\eta_c$ are both
$\bar
p=[(\sqrt{P^2+m_c^2}-P)/\sqrt{2},(\sqrt{P^2+m_c^2}+P)/\sqrt{2},\bm{0}_\perp]$,
where $P$ is the magnitude of the 3-momentum of any of the $Q$s or $\bar
Q$s. The momentum of the virtual photon is $Q=2(p+\bar p)$, which implies
that $Q^2=16(P^2+m_c^2)$. If a momentum $k$ has light-cone components
whose orders of magnitude are $P\lambda[1,(\eta^+)^2,\eta^+]$, then we
say that $k$ is soft if $\lambda\ll 1$, and we say that $k$ is collinear
to plus if $\eta^+\ll 1$. If $k$ has light-cone components whose orders
of magnitude are $P\lambda[(\eta^-)^2,1,\eta^-]$, then we say that $k$
is soft if $\lambda\ll 1$, and we say that $k$ is collinear to minus if
$\eta^-\ll 1$. Hence, $p$ is collinear to plus and $\bar p$ is collinear
to minus in the limit $m_c^2/P^2\to 0$.

\subsection{Evaluation of the diagrams}

\begin{figure}
\begin{center}
\includegraphics[width=1\textwidth]{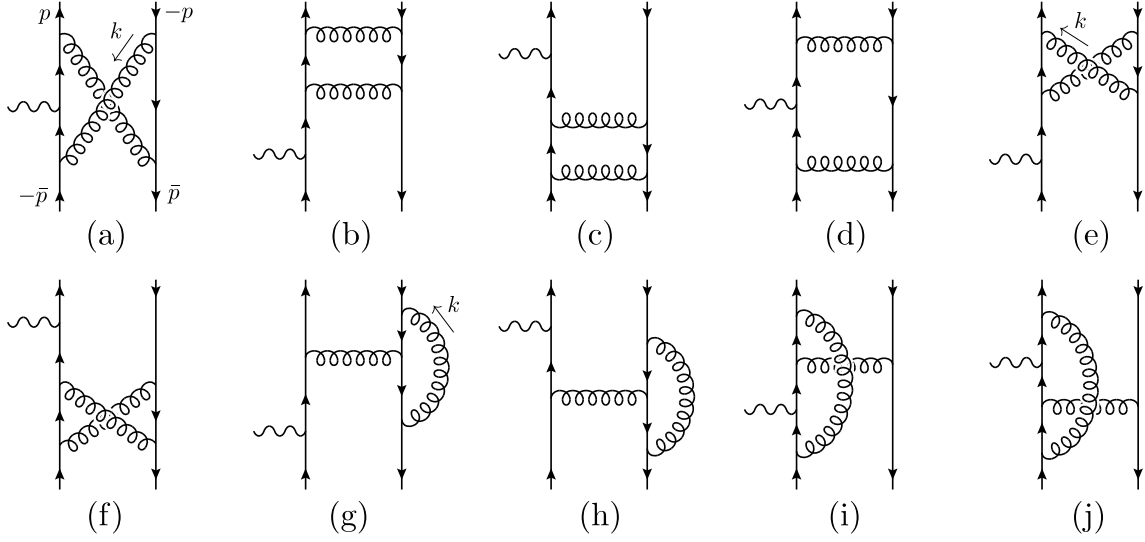}
\caption{
One-loop diagrams that produce double logarithms of $Q^2 /m_c^2$. 
The upper $Q\bar Q$ pair corresponds to the $J/\psi$, and the 
lower $Q\bar Q$ pair corresponds to the $\eta_c$.
}
\label{fig:oneloop}%
\end{center}
\end{figure}

Now we calculate the double logarithms that arise from
the Feynman diagrams that contribute to the NLO QCD corrections to the
amplitude. The amplitudes for each diagram contain spin and color
projectors that put the $Q\bar Q$ pairs into states of definite spin and
color~\cite{Bodwin:2002hg}.
When the relative momentum of the $Q$ and $\bar Q$ in each charmonium is
zero, the spin-singlet and spin-triplet projectors are given by
\begin{eqnarray}
&&
\Pi_1 (\bar p, \bar p) =- \frac{1}{2 \sqrt{2} m} \gamma^5 (\bar p \!\!\!\!\!/ +
m_c),\\
&&
\Pi_3 (p,p,\lambda) =- \frac{1}{2 \sqrt{2} m} \epsilon \!\!\!\!\!/^* (\lambda)
(p\!\!\!\!\!/ + m_c),
\end{eqnarray}
where $\epsilon^* (\lambda)$ is the polarization vector for the 
$Q\bar Q$ pair in the spin-triplet state.

The NLO diagrams that contain the double logarithms in $Q^2/m_c^2$ are
shown in Fig.~\ref{fig:oneloop}. Because the process $e^+ e^- \to J/\psi
+ \eta_c$ does not satisfy quark-helicity conservation, its amplitude
must contain a numerator factor $m_c$, which produces a helicity flip.
Hence, the amplitude is suppressed by a factor of $m_c/Q$ relative to a
helicity-conserving amplitude. The factor $m_c$ can come from the
numerators of the quark propagators or from the numerators of the spin
projectors. We must also retain a nonzero quark mass in denominators
because logarithmic collinear and endpoint divergences that appear in
the calculation are sensitive to that mass.

A straightforward calculation of the diagram in
Fig.~\ref{fig:oneloop}$\rm (a)$ gives
\begin{eqnarray}
\label{eq:oneloopexample}%
&& 
i {\cal A}_{\rm LO} \times \frac{-i \alpha_s \pi Q^2}{2} 
\left(C_F - \frac{1}{2} C_A \right) 
\times \bigg \{
\int \frac{d^dk}{(2 \pi)^d} \frac{1}{(k^2+i \varepsilon) 
[ (k+p)^2 - m_c^2 + i \varepsilon] 
[ (k-\bar p)^2 -m_c^2 +i \varepsilon] }
\nonumber \\ && \hspace{15ex}
+ 
\int \frac{d^dk}{(2 \pi)^d} \frac{1}{[(k+p+\bar p)^2+i \varepsilon] 
[ (k+2 p+\bar p)^2 - m_c^2 + i \varepsilon] 
[ (k+p)^2 -m_c^2 +i \varepsilon] }
\nonumber \\ && \hspace{15ex}
+
\int \frac{d^dk}{(2 \pi)^d} \frac{1}{(k^2+i \varepsilon) 
[ (k+p)^2 - m_c^2 + i \varepsilon] 
[ (k+p+\bar p)^2 +i \varepsilon] }
+ \cdots
\bigg \},
\end{eqnarray}
where we retain only the terms that contain the double logarithms in 
$Q^2/m_c^2$. Here, $C_F = (N_c^2 -1)/(2 N_c)$, $C_A = N_c$, $d=4-2
\epsilon$, and 
$i {\cal A}_{\rm LO}$ is the LO (order-$\alpha_s$) contribution to
the amplitude 
$i {\cal A} [\gamma^* \to Q \bar Q_1 ({}^3S_1) + Q \bar Q_1 
({}^1S_0)]$:
\begin{equation}
i {\cal A}_{\rm LO} = \frac{-i 256 \pi \alpha_s C_F}{m_c Q^4} 
\epsilon^{\mu \nu \alpha \beta} 
\epsilon^*_\nu (\lambda) p_\alpha \bar p_\beta,
\end{equation}
where $\mu$ is the polarization of the virtual photon. 
We use the nonrelativistic normalization for the
spinors. $\epsilon^{\mu \nu \alpha \beta}$ is the totally antisymmetric
tensor in $4$ dimensions, for which we use the convention
$\epsilon_{0123} = 1$. We evaluate the first integral in
Eq.~(\ref{eq:oneloopexample}) using dimensional regularization to control 
the soft divergence. We find that
\begin{eqnarray}
{\cal S} &\equiv& 
\int \frac{d^dk}{(2 \pi)^d} \frac{1}{(k^2+i \varepsilon) 
[ (k+p)^2 - m_c^2 + i \varepsilon] 
[ (k-\bar p)^2 -m_c^2 +i \varepsilon] }
\nonumber \\ 
&=& \frac{i}{4 \pi^2 Q^2} 
\bigg\{ \left[ \frac{1}{\epsilon_{\rm IR}} - \log (m_c^2/\mu^2) \right] 
\log (m_c^2/Q^2) + \frac{1}{2} \log^2 (m_c^2/Q^2) + \cdots \bigg\}, 
\end{eqnarray}
where we retain only the terms that are singular in $\varepsilon$ and the
double logarithms. 

We have carried out a detailed analysis of the first integral in
Eq.~(\ref{eq:oneloopexample}) that makes use contour integration for the
integration over $k_0$. That analysis shows that double logarithm in
this integral comes from the region in which the gluon with momentum
$k$ is simultaneously soft and collinear.
That is, it is a Sudakov double
logarithm. The second integral in Eq.~(\ref{eq:oneloopexample})
is identical to the first integral if we
change the loop momentum to $k+p+\bar p$. Hence, the second integral
gives a Sudakov double logarithm that comes from the region in
which the gluon with the momentum $k+p+\bar p$ is simultaneously soft
and collinear. The last integral in Eq.~(\ref{eq:oneloopexample})
yields
\begin{equation}
{\cal E} \equiv \int \frac{d^4k}{(2 \pi)^4} \frac{1}{(k^2+i \varepsilon) 
[ (k+p)^2 - m_c^2 + i \varepsilon] 
[ (k+p+\bar p)^2 +i \varepsilon] }
= \frac{i}{8 \pi^2 Q^2} 
\left[ 
\log^2 (m_c^2/Q^2) + \cdots \right], 
\end{equation}
where again we show only the double-logarithmic contributions. (In this
integral, there is no singularity in $\varepsilon$.) A detailed
analysis, which makes use of contour integration for the $k_0$
integration, shows that the double logarithm in this integral comes from
the momentum region in which the gluons carry almost all of the
collinear-to-plus and collinear-to-minus  momenta from the
spectator-quark line to the active-quark line. We call a
contribution from this momentum region an endpoint contribution. An
important observation is that $\cal E$ can be made to look similar to
the integral that gives the Sudakov double logarithm by changing the
loop momentum to the spectator-quark momentum $\ell = -k-p$. Then,
we have
\begin{equation}
{\cal E} 
= 
\int \frac{d^4\ell}{(2 \pi)^4} \frac{1}{
( \ell^2 - m_c^2 + i \varepsilon) 
[ (\ell+p)^2 + i \varepsilon] 
[ (\ell-\bar p)^2 +i \varepsilon] 
}. 
\label{reinterpret}
\end{equation}
It follows that the endpoint double logarithm arises from the region
in which the momentum of the spectator quark, $\ell$, is soft and
collinear. From Eq.~(\ref{reinterpret}), we see that the soft and
collinear divergences are regulated by $m_c$. An analysis of
Eq.~(\ref{reinterpret}) also shows that single logarithms of
$Q^2/m_c^2$ can arise from the region in which the momentum of the
spectator quark is soft.

Carrying out similar analyses of the remaining diagrams that
contribute to the amplitude $\gamma^* \to Q \bar Q_1 ({}^3S_1) + Q \bar
Q_1 ({}^1S_0)$ at order $\alpha_s^2$, we find that the double logarithms
in each diagram are accounted for by Sudakov double logarithms and
endpoint double logarithms. The Sudakov and endpoint double logarithms
that arise from each diagram are summarized in
Table.~\ref{Table:Double_Logarithms}. Our result for the sum of double
logarithms in all of the NLO diagrams agrees with the results in
Refs.~\cite{Gong:2007db, Jia:2010fw}. We also find that our results in
the Feynman gauge for the double logarithm in each diagram agree with
the results that were obtained in carrying out the calculation of
Ref.~\cite{Jia:2010fw}.\footnote{We thank Yu Jia and Xiu-Ting Yang for
providing us with the logarithmic contributions of the individual NLO
Feynman diagrams.}

Our detailed calculations are consistent with two general properties
of the Sudakov and endpoint singular regions: (1) the Sudakov double
logarithms cancel in the sum over diagrams; (2) the endpoint region
produces only logarithmic singularities, not power singularities. In the
following sections, we show how these general properties arise from
soft-collinear approximations that are valid in the Sudakov and endpoint
regions.

\begin{table} 
\begin{center}
\begin{tabular}{|c|c|c|c|} 
\hline
Diagram & Endpoint double logarithm & Sudakov double logarithm 
\\
\hline 
(a) & $(C_F-\frac{1}{2} C_A) {\cal E}$ & $2 (C_F-\frac{1}{2} C_A) {\cal S}$  
\\
(b) & $\phantom{1}\; C_F {\cal E}$ & $0$  
\\
(c) & $2\; C_F {\cal E}$ & $0$  
\\
(d) & $\frac{1}{2}\; C_F {\cal E}$ & $0$  
\\
(e) & $(C_F - \frac{1}{2} C_A){\cal E}$ & 
$\phantom{-} (C_F - \frac{1}{2} C_A) {\cal S}$  
\\
(f) & $0$ & $\phantom{-}(C_F - \frac{1}{2} C_A) {\cal S}$  
\\
(g) & $0$ & $-(C_F - \frac{1}{2} C_A) {\cal S}$  
\\
(h) & $0$ & $-(C_F - \frac{1}{2} C_A) {\cal S}$  
\\
(i) & $0$ & $-(C_F - \frac{1}{2} C_A) {\cal S}$  
\\
(j) & $0$ & $-(C_F - \frac{1}{2} C_A) {\cal S}$  
\\
\hline
\end{tabular} 
\end{center}
\caption{Endpoint and Sudakov double logarithms in each diagram, 
in units of $i {\cal A}_{\rm LO} \times (-i \alpha_s \pi Q^2)/2$. 
} 
\label{Table:Double_Logarithms} 
\end{table}

\section{General analysis of the Sudakov double 
logarithms\label{sec:sudakov}}

As we have mentioned, Sudakov double logarithms arise from a region in
which the momentum of a gluon is simultaneously soft and collinear. We
would like to apply a collinear approximation to such gluons. Consider,
for example, Figs.~\ref{fig:oneloop}$\rm (e)$ and $\rm (g)$, in
which a gluon with momentum $k$ that is collinear to minus is emitted
from a quark line with momentum $\bar p$. Then, the quark-gluon vertex
and the two propagator numerators surrounding it can be written as
\begin{equation}
(\bar{p}\sl +m_c)
\gamma^\mu
( \bar{p}\sl + k\sl +m_c)
=2(\bar{p}^\mu + k^\mu)(\bar{p}\sl 
+m_c)-k\sl\gamma^\mu  m_c,
\label{collinear-to-minus}
\end{equation}
where $\mu$ is the polarization index of the gluon, we have used 
$k\sl \bar{p}\sl \propto \bar{p}\sl \bar{p}\sl$ and 
$\bar{p}\sl \bar{p}\sl =m_c^2$, and we have dropped 
terms of order $m_c^2$. In the collinear-to-minus approximation, one 
retains only the first of the two terms on the right side of 
Eq.~(\ref{collinear-to-minus}). In the case of nonzero quark masses, 
this approximation is not valid in general. However, if $k$ is soft in 
comparison to $\bar{p}$, as 
well as collinear, then we can drop the second term on the right side of
Eq.~(\ref{collinear-to-minus}), and the standard collinear approximation 
holds. [We can also drop $k$ in the first term on the right side of
Eq.~(\ref{collinear-to-minus}).] Since the current in 
Eq.~(\ref{collinear-to-minus}) now lies in the minus light-cone
direction, up to terms of order $m_c^2$, we can make a
collinear-to-minus approximation in the gluon
propagator~\cite{Bodwin:1984hc, Collins:1985ue, Collins:1989gx}, by
making the replacement in the polarization tensor
\begin{equation}
g_{\mu\nu}\to \frac{k_\mu p_\nu}{k\cdot p- i\varepsilon},
\label{coll-prop}
\end{equation}
where the index $\nu$ corresponds to the attachment of the gluon to the
quark line that is collinear to minus and the sign of $i\varepsilon$ is
fixed by the sign in the original Feynman diagram. This approximation is
valid unless the $\mu$ attachment of the gluon is to a line that is also
collinear to minus. Hence, the approximation always holds for the
diagrams that produce Sudakov logarithms because the invariant $Q^2$ in
the logarithm can appear only if the soft-collinear gluon connects a
line carrying momentum $\bar{p}$ with a line carrying momentum $p$. The
replacement (\ref{coll-prop}) can also be regarded as a soft
approximation~\cite{Grammer:1973db, Collins:1981uk} to the $\mu$
attachment of the gluon. However, the collinear approximation can be
more useful in applications other than the present one because its form
is independent of the direction of the momentum $p$, while the form of
the soft approximation is not.
  
For the diagram of Fig.~\ref{fig:oneloop}$\rm (e)$ we can apply the
soft-collinear approximation (\ref{coll-prop}), where the index $\mu$
corresponds to the connection of the collinear-to-minus gluon to the
active-quark line that carries momentum $p$:
\begin{equation}
\bar{u} (p) \gamma^\nu \frac{1}{p \!\!\!\!\!/ - k \!\!\!\!\!/ -m_c
+i \varepsilon}
\to 
\bar{u} (p) \frac{k\sl p^\nu}{k \cdot p-i\varepsilon}  
\frac{1}{p \!\!\!\!\!/ - k \!\!\!\!\!/ -m_c
+i \varepsilon}
=
- \bar{u} (p) \frac{p^\nu}{k \cdot p-i\varepsilon},
\end{equation}
where, in the last equality, we have applied the graphical Ward
identity (Feynman identity). Similarly, for the diagram of
Fig.~\ref{fig:oneloop}$\rm (g)$ we can apply the soft-collinear
approximation (\ref{coll-prop}), where the index $\mu$ corresponds to
the connection of the collinear-to-minus gluon to the spectator-quark
line that carries momentum $p$:
\begin{eqnarray}
\frac{1}{-p \!\!\!\!\!/ + k \!\!\!\!\!/ -m_c+i \varepsilon}
\gamma^\nu v(p) 
&\to& 
\frac{1}{-p \!\!\!\!\!/ + k \!\!\!\!\!/ -m_c+i \varepsilon}
\frac{k\sl p^\nu}{k \cdot p-i\varepsilon} 
v(p)
= + \frac{p^\nu}{k \cdot p-i\varepsilon} v(p). 
\end{eqnarray}
Because the outgoing $Q\bar Q$ pairs are in color-singlet states,
these two contributions have the same color factor and cancel. This type
of cancellation extends to all diagrams involving a gluon with
soft-collinear-to-minus momentum. Similar cancellations occur for
diagrams involving a gluon with soft-collinear-to-plus momentum.
Therefore, in the sum of all diagrams, Sudakov double logarithms cancel.

\section{General analysis of the endpoint region \label{sec:endpoint}}

As we have mentioned, the endpoint double logarithms of $Q^2/m_{c}^2$
arise from the region of loop integration in which the momentum
$\ell$ of the internal spectator-quark line is simultaneously soft and
collinear. Hence, we need to consider only diagrams that can produce
such a momentum configuration. These diagrams are shown in
Figs.~\ref{fig:oneloop}$\rm (a)$--$\rm (f)$. (A diagram and its charge
conjugate give equal contributions to the amplitude. The
charge-conjugate diagrams are not shown in Fig.~\ref{fig:oneloop}.) The
endpoint double logarithms arise from contributions in which there is a
numerator factor $m_c$, which produces the helicity flip,  and in which
integrals diverge logarithmically in the limit $m_c\to 0$. In general,
integrals can also diverge as inverse powers of $m_c$ in the limit
$m_c\to 0$, but, as we shall see, such contributions vanish when the
numerator trace is taken.

In diagrams $\rm (a)$, $\rm (e)$, and $\rm (f)$, the momenta of the
propagators on the active-quark lines contain both $p$ and $\bar p$.
Since $p\cdot \bar p\sim P^2\sim Q^2$, we can ignore $\ell$ in the
denominators of those propagators. In the limit $m_c\to 0$, the two
gluon-propagator denominators and spectator-quark-propagator denominator
produce factors $1/(\ell^2+2p\cdot \ell)$, $1/(\ell^2-2\bar p\cdot
\ell)$, and $1/\ell^2$, respectively, where we have dropped the $m_c^2$
terms in the propagator denominators. Hence, in order to obtain a
logarithmically divergent soft power count ($\lambda^{-4}$) and
logarithmically divergent collinear power counts [$(\eta^\pm)^{-4}$], we
must drop all numerator factors of $\ell$. This implies that the
helicity flip comes from the factor $m_c$ in the numerator of the
spectator-quark propagator.

In diagram (b), the momentum of the outermost active-quark propagator 
contains the momentum $p$, but not the momentum $\bar p$. Hence, in
the limit $m_c\to 0$, the denominator of this propagator produces a factor
$1/(\ell^2+4p\cdot \ell)$. Then, all of the propagator denominators
taken together produce a linearly divergent soft power count and a
linearly divergent collinear-to-plus power count. However, it is easy to
see that numerator factors reduce both of these power counts to
logarithmic ones. First, we rewrite the numerator factors that
are associated with the outermost gluon and the spin projector for the
$J/\psi$ $Q\bar Q$ pair as $\gamma_\mu (p\sl
-m_c){\epsilon\sl}^*\gamma^\mu=2m_c{\epsilon\sl}^*$, where we have used
the fact that $p\cdot \epsilon^*=0$. Now, because the numerator
power of $m_c$ is in this factor, the numerator of the spectator-quark
propagator must contribute a factor $\ell\sl$. Furthermore, the
numerator vanishes, up to terms of order $m_c^2$, if $\ell$ is proportional
to $p$ because there are then two factors of $p\sl$ that are separated
by $\gamma$ matrices with which they anticommute. Hence, in the trace
over the $\gamma$ matrices, $\ell$ must appear in the combination
$\ell\cdot p$. This numerator factor reduces both the soft and collinear
power counts to logarithmic ones.

In the case of diagram (c), the denominator of the outermost
active-quark propagator contributes a factor $1/(\ell^2-4\bar p\cdot
\ell)$. Hence, the soft and collinear-to-minus power counts from the
propagator denominators are both linearly divergent. Now, we rewrite
the numerator factors that are associated with the outermost gluon and
the spin projector for the $\eta_c$ $Q\bar Q$ pair as $\gamma_\mu
(-{\bar p}\sl-m_c)\gamma_5\gamma^\mu =(-2{\bar p}\sl +4m_c)\gamma_5$. If
the numerator factor $m_c$ comes from the spectator-quark propagator,
then there must be a factor $\ell\sl$ from the outermost active-quark
propagator, or else there are two adjacent factors of ${\bar p}\sl$ and
the numerator vanishes, up to terms of order $m_c^2$. If the numerator
factor $m_c$ does not come from the spectator-quark propagator, then
that propagator must yield a numerator factor $\ell\sl$. In both cases,
the numerator factor vanishes, up to terms of order $m_c^2$, if $\ell$
is proportional to $\bar p$ because, in that case, there are two factors
of ${\bar p}\sl$ that are either adjacent or are separated by $\gamma$
matrices with which they anticommute. Therefore, we conclude that the
trace contains a factor $\ell\cdot \bar p$, which reduces both the soft
and collinear-to-minus power counts to logarithmic ones.

In the case of diagram (d), the denominators of the active-quark
propagators produce factors $1/(\ell^2+4p\cdot \ell)$ and
$1/(\ell^2-4\bar p\cdot \ell)$. Hence, the propagator denominators, taken
together, produce a quadratically divergent soft power count and
linearly divergent collinear-to-plus and collinear-to-minus power
counts. One can apply the arguments that were used for the numerators of
diagrams (b) and (c) separately to each of the gluons in diagram (d),
with the conclusion that the numerator contains two factors of $\ell\sl$
and that the numerator vanishes, up to terms of order $m_c^2$, if $\ell$
is proportional to $p$ or to $\bar p$. Hence, the trace contains a
factor $\ell\cdot p\, \ell\cdot \bar p$ or a factor $\ell^2$. In either
case, the numerator factor reduces the soft and collinear power counts
to logarithmic ones.

In the case of one-loop corrections to helicity-conserving
charmonium-production processes, in which there is no numerator factor
$m_c$, the previous arguments show that the contribution from the region
in which the spectator quark carries a soft-collinear momentum vanishes,
implying that there are no endpoint double logarithms. This general
conclusion is confirmed in explicit calculations
\cite{Jia:2010fw,Dong:2011fb,Jia:2008ep}.

\section{Summary}
\label{sec:summary}%

In this work we have investigated the origin of the double logarithms
of $Q^2/m_c^2$ that appear in the NLO QCD corrections to the process $e^+
e^- \to J/\psi+\eta_c$. We find that the double logarithms in each
diagram are accounted for by Sudakov double logarithms and endpoint
double logarithms. The Sudakov double logarithms cancel in the sum of
all diagrams. We have given a general argument for this cancellation
that is based on the soft-collinear approximation and graphical Ward
identities. We have reinterpreted the region of a loop integration that
gives rise to an endpoint double logarithm as a leading region in which
the momentum of the spectator-quark line is both soft and collinear. Under
this reinterpretation, we would also expect single logarithms of
$Q^2/m_c^2$ to arise from a leading region in which the momentum of the
spectator-quark line is soft. This reinterpretation may be useful in
establishing an all-orders factorization theorem for helicity-flip
quarkonium production. Such a factorization theorem might allow one to
resum logarithms of $Q^2/m_c^2$ to all orders in $\alpha_s$ for
helicity-flip processes. We have also given a power-counting analysis of
the one-loop endpoint contributions, which shows that they can arise
only in the presence of a helicity flip and that the loop integration
can produce logarithms, but not inverse powers, of the heavy-quark mass
$m_c$.


\end{document}